\documentclass[prd,preprint,showpacs,nofootinbib]{revtex4}
\usepackage{graphicx}
\usepackage{amssymb}
\usepackage{amsmath}

\begin{document}

\begin{flushright}
\end{flushright}

\vskip4cm

\title{Baryon Asymmetry in a Heavy Moduli Scenario}
\author{Masahiro Kawasaki}
\author{Kazunori Nakayama}
\affiliation{Institute for Cosmic Ray Research,
University of Tokyo,
Kashiwa 277-8582, Japan}
\date{\today}

\begin{abstract}
In some models of supersymmetry breaking, modulus fields are heavy
enough to decay before BBN.  But the large entropy produced via moduli
decay significantly dilutes the preexisting baryon asymmetry of the
universe.  We study whether Affleck-Dine mechanism can provide enough
baryon asymmetry which survives the dilution, and find several
situations in which desirable amount of baryon number remains after the
dilution.  The possibility of non-thermal dark matter is also discussed.
This provides the realistic cosmological scenario with heavy moduli.
\end{abstract}

\maketitle

\section{Introduction}

Recent cosmological observations have revealed that the ordinary matter
contributes to the only small fraction of the energy density of the
universe, $\Omega_b h^2 \sim 0.022$ in terms of the density parameter,
and the remainder comes from ``dark'' components, dark matter and dark
energy, whose contributions are represented by $\Omega_m h^2 \sim 0.13$
and $\Omega_\Lambda \sim 0.72$ respectively \cite{Spergel:2006hy}.  The
existence of these dark components indicates physics beyond the standard
model such as supersymmetry (SUSY) \cite{Nilles:1983ge}, but on the
other hand, the non-standard physics is also considered to be imprinted
in the ordinary matter (baryon) component.  From the measurement of
cosmic microwave background (CMB) anisotropy and light element abundances
predicted by big-bang nucleosynthesis (BBN) \cite{Olive:1999ij}, it is
known that the baryon density of the universe is almost made only of
baryons, and anti-baryons do not exist .  But such observed amount of
baryon asymmetry can not be generated within the framework of the
standard model.  Thus, if there is an underlying physics beyond the
standard model, the baryon density of the universe $\Omega_b h^2 \sim
0.022$ should also be explained by violation of the baryon number and CP
built in the new physics, as well as the dark matter.  In supersymmetric
theory, which is one of the best motivated physics beyond the standard
model, there is an interesting mechanism to create baryon asymmetry.  In
SUSY there exist many scalar fields as superpartner of the standard
model fermion which carry baryon or lepton number.  Along some directions
of the configuration of these scalar fields, the scalar potential is
flat. Thus scalar fields corresponding to these flat directions can develop
to large field value during inflation, and subsequent evolution of the
scalar fields naturally leads to baryon asymmetry.  This is called
Affleck-Dine mechanism \cite{Affleck:1984fy}, which we will focus on
this paper.

On the other hand, global supersymmetry is naturally extended to the
local supersymmetry, which inevitably includes gravity, that is
supergravity.  In supergravity there appear long-lived massive particles
whose lifetime are typically longer than 1 sec and they decay after BBN
starts.  One is the gravitino, the superpartner of the graviton.  Gravitinos
are produced in high-temperature plasma via scatterings of the particles
in thermal bath and their subsequent decay may greatly affect the
standard cosmology \cite{Khlopov:1984pf,Bolz:2000fu}, or may overclose
the universe if they are stable \cite{Moroi:1993mb}.  Another is the
Polonyi field, which is a singlet scalar field introduced in order to
give the SUSY breaking masses to the superparticles, especially 
the mass of gauginos, in many models of SUSY breaking.
Generally the Polonyi field has the
large field value during inflation, and it begins to oscillate
coherently with initial amplitude of order reduced Planck scale $M_P$
when the Hubble parameter $H$ becomes equal to the gravitino mass $m_{3/2}$.  It has
extremely large energy density and its decay after BBN has catastrophic
effects on the standard cosmology~\cite{Coughlan:1983ci}.  Furthermore,
supergravity may be the low energy effective theory of string theory,
which is defined in 10 dimensional space-time.  In compactification of
such extra dimensions, there appear light scalar fields called moduli.
Generally moduli have the mass of order $m_{3/2}$ through non-perturbative
dynamics associated with SUSY breaking.  The dynamics of moduli fields
and cosmological difficulty they cause are similar to those of the
Polonyi field, and we call them moduli problem \cite{Banks:1993en}.

There are some suggestions to solve the moduli problem.  One possible
way is to use late-time inflation and the subsequent entropy production
in order to dilute the moduli abundance to cosmologically safe value.
Such late-time inflation is realized by thermal inflation models and
concrete examples for are found in
Refs.~\cite{Yamamoto:1985rd,Lyth:1995hj,Asaka:1999xd}.  The other way is
to make moduli heavy enough to decay well before BBN.  The modulus mass
larger than about 100 TeV is safe and such large mass is naturally
realized in anomaly-mediated SUSY breaking models, where other SUSY
particles obtain the mass of order $\sim m_{3/2}/(8\pi^2) \sim 1$
TeV~\cite{Randall:1998uk}.

But the scenario is not complete.  In late-time entropy production
scenario, the preexisting baryon asymmetry is also diluted.  The
reheating temperature after thermal inflation is typically less than a
few GeV, and hence almost all baryogenesis mechanisms which rely on high
energy physics do not work.  The variant type of Affleck-Dine mechanism
after thermal inflation may work \cite{Stewart:1996ai} and to the best
of our knowledge it is only possibility to create enough baryon
asymmetry in the presence of thermal inflation.  In heavy moduli
scenario, significant entropy released by the decay of moduli also
dilute the preexisting baryon asymmetry.  In previous works
\cite{Moroi:1994rs,Moroi:1999zb}, it was assumed that Affleck-Dine
mechanism can create enough baryon asymmetry which survives the dilution
from moduli decay.  However, in fact, for large $m_{3/2}$ ordinary
Affleck-Dine mechanism does not work due to the non-trivial potential
minima of the Affleck-Dine field \cite{Fujii:2002kr,Kawasaki:2000ye}.

In this paper we study whether the sufficient baryon asymmetry is
created in heavy moduli scenario, such as anomaly-mediated SUSY breaking
models or mixed modulus anomaly mediation (or mirage mediation) models
\cite{Choi:2004sx}.  
Mirage mediation models are based on the concrete model of KKLT
flux compactification \cite{Kachru:2003aw} and the lightest modulus mass
is predicted as $m_{\chi}\sim (4\pi^2)m_{3/2}$.  
Thus in the following we consider the typical situations where the modulus mass $m_\chi$ 
is of the order $m_{3/2}$ and $(4\pi^2)m_{3/2}$
as reference values, although we do not specify the concrete SUSY breaking models.
The only assumption on which our analysis is based
is the hierarchical relation between the gravitino mass and other SUSY particle masses.
In both scenarios (anomaly- or mirage mediation models) other
SUSY particles have mass of order $\sim m_{3/2}/(8\pi^2)$ and hence the
gravitino is considered to be as heavy as 100 TeV.  As we will explain
later, Affleck-Dine mechanism for such large $m_{3/2}$ is highly
non-trivial.  One possible way to incorporate Affleck-Dine mechanism in
anomaly- or mirage-mediation model is to introduce the gauged
U(1)$_{B-L}$ symmetry \cite{Fujii:2001sn}.  But we pointed out that
Affleck-Dine mechanism with large $m_{3/2}$ works without any additional
assumption if the reheating temperature is relatively high
\cite{Kawasaki:2006yb}.  We examine these two scenarios and find
possible parameter region in which the desired amount of baryon
asymmetry is generated.

This paper is organized as follows.  In Sec.~\ref{sec:dynamics}, we
briefly review the cosmological dynamics of moduli fields.  In
Sec.~\ref{sec:earlyosc}, Affleck-Dine baryogenesis in high reheating
temperature scenario is discussed.  In Sec.~\ref{sec:gauged},
Affleck-Dine baryogenesis in gauged U(1)$_{B-L}$ model is discussed.
Somewhat similar to this case, but the case without superpotential and
Q-ball domination is discussed in Sec.~\ref{sec:W=0}.  The decay of
moduli may cause another difficulty, especially LSP overproduction from
the decay of moduli-induced gravitinos.  We give possible solutions to
this problem in Sec.~\ref{sec:remarks}.  We conclude in
Sec.~\ref{sec:conclusion}.

\section{Dynamics of modulus fields}   \label{sec:dynamics}

In general, a modulus field has the mass of order $m_{3/2}$ in the
presence of SUSY breaking.  In the early universe, the inflaton
dominates the energy density and this vacuum energy also breaks
supersymmetry.  As a result the modulus field obtains Hubble-induced
SUSY breaking mass of order $H$ \cite{Dine:1995uk}.  Thus, during
inflation the modulus has very large mass and sits at the origin.
However, this high-energy minimum does not need to coincide with the
low-energy true minimum of the potential.  In general, these two minima
are expected to be separated by the Planck scale and hence when $H$
becomes smaller than the modulus mass $m_\chi$, the modulus field begins
to oscillate with the initial amplitude $\chi_0 \sim M_P$.  This
coherent oscillation of the modulus field has the energy density $\sim
m_\chi^2 M_P^2$ initially, and the total energy density of the universe
is given by $\sim 3m_\chi ^2 M_P^2$.  We can see that the modulus has
inevitably large energy density comparable to the total energy density
and dominates the universe as soon as inflaton decays (in the case of
$m_\chi > \Gamma_I$, where $\Gamma_I$ denotes the decay rate of the
inflaton) or the modulus field begins to oscillate (in the case of
$\Gamma_I > m_\chi$).

In the case of $m_\chi > \Gamma_I$ where inflaton decays after the
moduli start to oscillate, the moduli-to-entropy ratio is given by
\begin{equation}
\begin{split}
  \frac{\rho_\chi}{s}=&\frac{m_\chi^2 \chi(T_R)^2}{4\rho(T_R)/3T_R}
  =\frac{1}{4}T_R\left (\frac{\chi_0}{M_P} \right )^2 \\
  &\sim 2.5\times 10^4 \mathrm{GeV}
  \left (\frac{T_R}{10^5 \mathrm{GeV}} \right )
	\left (\frac{\chi_0}{M_P} \right )^2,   \label{modBB2}
\end{split}	
\end{equation}
where $T_R$ denotes the reheating temperature from inflaton decay.
On the other hand, in the case of $m_\chi < \Gamma_I$, it is given by
\begin{equation}
\begin{split}
   \frac{\rho_\chi}{s}=& \frac{1}{4}
   \left (\frac{90}{\pi^2 g_*} \right )^{1/4}
   m_\chi^{1/2}M_P^{1/2}\left (\frac{\chi_0}{M_P} \right )^2 \\
   &\sim1.2\times 10^{11} \mathrm{GeV}
   \left (\frac{90}{\pi^2 g_*} \right )^{1/4}
   \left (\frac{m_\chi}{100 \mathrm{TeV}} \right )^{1/2}
   \left (\frac{\chi_0}{M_P} \right )^2.  \label{modBB1}
\end{split}
\end{equation}
In both cases the moduli abundance largely exceeds the critical density
of the universe, $\rho_c/s_0 \sim 3.6 \times 10^{-9}h^2$ GeV.  If we
parametrize the decay width of the modulus as
\begin{equation}
   \Gamma_\chi = \frac{c}{4\pi}\frac{m_\chi^3}{M_P^2},
\end{equation}
the decay temperature of moduli $T_\chi$ is given by
\begin{equation}
	T_\chi \sim 5.5 \mathrm{MeV} \sqrt{c} 
	\left ( \frac{m_\chi}{100 \mathrm{TeV}} \right )^{3/2}.
	\label{modbaryon:T_chi}
\end{equation}
Hence, the decay temperature of modulus typically takes a value from a
few MeV to a few GeV for $100$ TeV $\lesssim m_\chi \lesssim (4\pi^2)
100$ TeV.  The late decay of moduli with such large abundance has
significant effects on BBN \cite{Kawasaki:1994af}, CMB
\cite{Fixsen:1996nj} and diffuse X$(\gamma)$-ray background
\cite{Kawasaki:1997ah}, which results in strong disagreement with
observations.  But for the modulus mass larger than about $100$ TeV, the
moduli decay before BBN and do not spoil the success of standard BBN.
The universe is finally reheated by the moduli with reheating
temperature $T_\chi$ \footnote{If the modulus field does not have the
Hubble mass and obtain unsuppressed quantum fluctuation during
inflation, it can be the interesting candidate of the curvaton
\cite{Moroi:2001ct,Hamaguchi:2003dc}.  But we do not go into the details
of this issue.}.

\section{Affleck-Dine baryogenesis with early oscillation}  
\label{sec:earlyosc}

\subsection{The model}

In minimal supersymmetric standard model (MSSM), there exist some
configurations of the scalar fields (flat directions) along which
scalar potential vanishes in supersymmetric limit within the
renormalizable terms \cite{Gherghetta:1995dv}.  A flat direction is
parameterized by the single complex scalar $\phi$, which we call
Affleck-Dine field.  The Affleck-Dine field feels potentials from
non-renormalizable superpotentials represented as
\begin{equation}
	W_{\mathrm{NR}}=\frac{\phi^n}{nM^{n-3}}
\end{equation}
where $M$ is the effective cutoff scale and $n\geq 4$.  Including SUSY
breaking effects, the potential for the Affleck-Dine field is written as
\begin{equation}
    V_S(\phi)= m_{\phi}^2|\phi|^2 +
     \left ( a_m m_{3/2 }
      \frac{\phi^n}{nM^{n-3}} + \mathrm{h.c.} \right )
     +\frac{|\phi|^{2(n-1)}}{M^{2(n-3)}},  \label{zeropot}
\end{equation}
where $a_m$ is $O(1)$ numerical coefficient.  There are other sources
for the scalar potentials.  As explained in sec.\ref{sec:dynamics}, the
scalar fields obtain Hubble-induced SUSY breaking terms such as
\begin{equation}
   V_H(\phi) = -c_H H^2 |\phi|^2 
    +\left ( a_H H\frac{\phi^n}{nM^{n-3}} + \mathrm{h.c.} \right ),
\end{equation}
where $c_H$ and $a_H$ are $O(1)$ coefficients~\footnote{In some
inflation models such as $D$-term inflation \cite{Halyo:1996pp},
Hubble-induced term does not arise during
inflation \cite{Kolda:1998kc}.  }. Here we assume $c_H >0$. Furthermore,
in high-temperature environment of the early universe, thermal
corrections to the scalar potential also arise.  These are
\cite{Allahverdi:2000zd,Anisimov:2000wx}
\begin{equation}
   V_T(\phi)=\sum _{f_k|\phi |<T}c_k f_k^2 T^2|\phi |^2
    + a\alpha(T)^2 T^4 
    \log \left(  \frac{|\phi |^2}{T^2}  \right)
\end{equation}
where $c_k$ is a constant of order unity, $f_k$ denotes gauge or Yukawa
couplings relevant for the Affleck-Dine field, and $a$ is a constant of
order unity assumed to be positive, which is determined by the two-loop
finite-temperature effective potential for the Affleck-Dine field.  Then
the total scalar potential for the Affleck-Dine field is sum of them,
\begin{equation}
	V(\phi) = V_S(\phi)+V_H(\phi)+V_T(\phi).
\end{equation}

Let us summarize the dynamics of Affleck-Dine field.
First, it is trapped by the minimum determined by the balance of 
negative Hubble-induced mass term and non-renormalizable term,
\begin{equation}
	|\phi | \simeq (HM^{n-3})^{1/(n-2)},
	\label{minimum}
\end{equation}
and tracks this minimum as $H$ becomes small.
The important fact is that without the finite-temperature effect, 
the scalar potential has the global minimum at
\begin{equation}
	|\phi |_{\mathrm{min}}\sim 
	\left(\frac{|a_m|}{n-1} m_{3/2}M^{n-3} \right)^{1/(n-2)}. 
	\label{breakmin}
\end{equation}
if $m_{3/2} $ is much greater than $m_\phi$, which is the situation we
are interested in.  Thus the Affleck-Dine field is eventually trapped by
this minimum and leads to charge or color breaking vacuum, which is a
disaster.  But finite-temperature effects can save the situation.
Including finite-temperature effects, when $H$ becomes equal to
$H_{\mathrm{os}}$ determined by
\begin{equation}
	H_{\mathrm{os}}^2\sim m_\phi^2 + 
	\sum _{f_k|\phi |<T}c_k f_k^2 T^2 
	+ a\alpha(T)^2 \frac{T^4}{|\phi |^2},
	\label{osc.condition}
\end{equation}
the Affleck-Dine field begins to oscillate around its minimum of the
potential.  The important fact is that if the thermal log term
dominates the potential and oscillation begins by this term, the
Affleck-Dine field will be taken to the origin without trapped by the
global minimum~\cite{Kawasaki:2006yb}.  Through the process of field
evolution, the Affleck-Dine field receives angular kick from $A$-terms
and results in elliptical motion around the origin.  Hence the baryon
number is generated and conserved in comoving volume.  In fact, as we
will see, for high reheating temperature from inflaton and high field
value, the oscillation starts when thermal logarithmic term dominates
the potential and Affleck-Dine mechanism works well.  Note that although
the resultant vacuum is meta-stable, the lifetime of the false vacuum is
much longer than the age of the universe \cite{Kawasaki:2000ye}.

\subsection{Baryon asymmetry}

Next we estimate the baryon asymmetry in the presence of a heavy modulus
field.  In the case of early oscillation due to thermal logarithmic
potential, $H_{\mathrm{os}}$ is given by
\begin{gather}
   H_{\mathrm{os}} = \alpha T_R \left ( \frac{M_P}{M} \right )^{1/2}
     ~~~~~\mathrm{for}~~~n=4,\\
   H_{\mathrm{os}}= \left (\alpha^2 T_R^2 M_P M^{-3/2} \right )^{2/3}
	~~~~~\mathrm{for}~~~n=6.    \label{modbaryon:Hos}
\end{gather}
Hereafter we consider $n=4$ and $n=6$ case only, because flat directions
with $n \geq 7$ are lifted by the superpotential of the form $\psi
\phi^{n-1}/M^{n-3}$ where $\psi$ represents scalar field other than
Affleck-Dine field and can not generate baryon asymmetry via
Affleck-Dine mechanism \cite{Gherghetta:1995dv}.  We need following
constraints for this scenario to work.  One is the condition that early
oscillation occurs ($H_{\rm os}> m_{\phi}$), this leads to
\begin{equation}
 \begin{split}
   T_R & \gtrsim \frac{m_\phi}{\alpha}
      \left ( \frac{M}{M_P} \right )^{1/2} \\
	& \sim 1\times 10^3 \mathrm{GeV} 
	\left( \frac{0.1}{\alpha}\right) 
       \left ( \frac{m_\phi}{100 \mathrm{GeV}} \right )
	\left ( \frac{M}{M_P} \right )^{1/2}
 \end{split}
\end{equation}
for $n=4$, and
\begin{equation}
\begin{split}
	T_R  ~&\gtrsim ~ \frac{1}{\alpha} 
	\left( \frac{m_\phi^3 M^3}{M_P^2} \right)^{1/4}  \\
	 &\sim  3\times 10^5 \mathrm{GeV}
	\left( \frac{0.1}{\alpha}\right) 
	\left( \frac{m_\phi}{100 \mathrm{GeV}} \right)^{3/4}
	\left( \frac{M}{10^{15} \mathrm{GeV}} \right)^{3/4}
\end{split}
\end{equation}
for $n=6$.  The other is the condition that thermal correction hide the
valley of the potential around the true charge breaking minimum and this
thermal logarithmic potential leads Affleck-Dine field to the origin.
This condition is written in the form $\alpha^2 T_{\mathrm{os}}^4
\gtrsim | V(|\phi|_\mathrm{min}) |$, explicitly,
\begin{equation}
 \begin{split}
   T_R &\gtrsim \alpha^{-1}m_{3/2} 
       \left ( \frac{M_P}{M} \right )^{1/2} \\
   &\sim 1\times 10^6 \mathrm{GeV} \left ( \frac{0.1}{\alpha} \right )
	\left( \frac{m_{3/2}}{100 \mathrm{TeV}} \right)
	\left ( \frac{M}{M_P} \right )^{1/2}  
	\label{modbaryon:TR(n=4)}
 \end{split}
\end{equation}
for $n=4$ and
\begin{eqnarray}
	T_R  & \gtrsim & \alpha ^{-1} 
	M^{3/4}M_P^{-1/2} m_{3/2}^{3/4}
	\nonumber \\ 
    &\sim &  8\times 10^5 \mathrm{GeV}
	\left( \frac{0.1}{\alpha}\right) 
	\left( \frac{m_{3/2}}{10 \mathrm{TeV}}\right)^{3/4}  
	\left( \frac{M}{10^{15} \mathrm{GeV}}\right)^{3/4}
	\label{modbaryon:TR(n=6)}
\end{eqnarray}
for $n=6$.  We can see that in general the latter condition is severer
whenever $m_{3/2}>m_\phi$, which is always satisfied in anomaly-mediated
SUSY breaking.

Now let us estimate the baryon asymmetry in the presence of modulus
fields.  It is convenient to express the baryon-to-entropy ratio as
\begin{equation}
   \frac{n_B}{s}=\frac{n_B}{\rho_\chi}\frac{\rho_\chi(T_\chi)}{s(T_\chi)}
    =\frac{n_B}{\rho_\chi}\frac{3T_\chi}{4}.
\end{equation}
The ratio $n_B/\rho_\chi$ is fixed when both the Affleck-Dine field and
modulus field begin to oscillate.  When $H_{\mathrm{os}}>m_\chi$, this
ratio is fixed at the onset of the oscillation of the moduli, $H =
m_\chi$.  On the other hand, when $H_{\mathrm{os}}<m_\chi$, the ratio is
fixed at the beginning of Affleck-Dine field oscillation,
$H=H_{\mathrm{os}}$.  From Eqs.(\ref{modbaryon:Hos}) and
(\ref{modbaryon:TR(n=4)}) or (\ref{modbaryon:TR(n=6)}),
\begin{equation}
	H_{\mathrm{os}} \gtrsim m_{3/2}
\end{equation}
must always hold.  Thus if we assume $m_\chi \sim m_{3/2}$ we can safely
focus on the case $H_{\mathrm{os}}>m_\chi$.  But in some models based on
string theory, $m_\chi \gg m_{3/2}$ might be possible.  In mirage
mediation model, the modulus mass is predicted as $m_\chi \sim
4\pi^2 m_{3/2}$ \cite{Choi:2004sx}.  Although in such a model
$H_{\mathrm{os}} < m_\chi$ is still possible, we mainly focus the case
$H_{\mathrm{os}}>m_{\chi}$  and
briefly discuss about the modification in the case
$H_{\mathrm{os}}<m_\chi$.  The condition $H_{\mathrm{os}}>m_\chi$ is
rewritten as
\begin{equation}
\begin{split}
	T_R &\gtrsim \alpha^{-1}m_{\chi} \left ( \frac{M_P}{M} \right )^{1/2} \\
	&\sim 1 \times 10^6\mathrm{GeV} \left ( \frac{0.1}{\alpha} \right )
	\left( \frac{m_{\chi}}{100\mathrm{TeV}} \right)
	\left ( \frac{M}{M_P} \right )^{1/2}   
	\label{modbaryon:Hos>m(n=4)}
\end{split}
\end{equation}
for $n=4$ and
\begin{equation}
\begin{split}
	T_R &\gtrsim \alpha^{-1}m_\chi^{3/4}M^{3/4}M_P^{-1/2}\\
	&\sim 2 \times 10^9\mathrm{GeV} 
	\left( \frac{0.1}{\alpha}\right) 
	\left( \frac{m_{\chi}}{100 \mathrm{TeV}}\right)^{3/4}
	\left( \frac{M}{M_P}\right)^{3/4}   .
	\label{modbaryon:Hos>m(n=6)}
\end{split}
\end{equation}
for $n=6$.  High reheating temperature from inflaton is not a problem as
far as the moduli decay well before BBN and non-thermal LSPs associated
with modulus decay do not overclose the universe.  In fact it is
possible that non-thermal LSPs from the decay of moduli account for the
present matter density of the universe (see Sec.~\ref{sec:remarks}).

\subsubsection{$H_{\mathrm{os}}>m_\chi$}

In this case the baryon-to-moduli ratio $n_B/\rho_\chi$ is fixed at
$H=m_\chi$ where the modulus field begins to oscillate with amplitude
$\chi_0 \sim M_P$,
\begin{equation}
	\frac{n_B}{\rho_\chi}=\frac{n_B(t_{\mathrm{os}})}{m_\chi^2 \chi_0^2}
	\left ( \frac{a(t_{\mathrm{os}})}{a(t_{\mathrm{mod}})} \right )^3,
	\label{modbaryon:n_B/rho_chi(i)}
\end{equation}
where $t_{\rm mod} \simeq m_{\chi}^{-1}$.  In order to get correct
estimation, we must specify the decay epoch of inflaton, whose decay
rate is denoted as $\Gamma_I$.  Thus depending on $\Gamma_I$, three
scenarios are available: (a) $\Gamma_I>H_{\mathrm{os}}>m_\chi$, (b)
$H_{\mathrm{os}}>\Gamma_I>m_\chi$, (c)
$H_{\mathrm{os}}>m_\chi>\Gamma_I$.  Note that in case (a), at the
beginning of oscillation of the Affleck-Dine field the universe already
enters radiation dominated era and estimation of baryon number is
somewhat different from the other two cases.  Before the estimation, we
see the conditions when the case (a), (b) and (c) are realized.  The
condition that $m_\chi>\Gamma_I$ can be written as
\begin{equation}
	T_R \lesssim 2\times 10^{11} \mathrm{GeV} 
	\left( \frac{m_{\chi}}{100 \mathrm{TeV}}\right)^{1/2}.
	\label{modbaryon:m>Gamma}
\end{equation}
On the other hand, the condition $H_{\mathrm{os}}
>\Gamma_I$ can be rewritten as follows,
\begin{equation}
\begin{split}
   &T_R \lesssim 5\times 10^{16} \mathrm{GeV} 
    \left( \frac{\alpha}{0.1} \right) 
    \left( \frac{M_P}{M}\right)^{1/2} ~~~~~\mathrm{for}~~~n=4,\\
   &T_R \lesssim 2\times 10^{15} \mathrm{GeV} 
    \left( \frac{\alpha}{0.1}\right) ^2
    \left( \frac{M_P}{M}\right)^{3/2} ~~~~~\mathrm{for}~~~n=6
	\label{modbaryon:Hos>Gamma}
\end{split}
\end{equation}
which is satisfied for natural range of parameters.  In other words,
unless reheating temperature is unnaturally high, case (a) is not
realized.  The conditions (\ref{modbaryon:Hos>m(n=4)}) (or
(\ref{modbaryon:Hos>m(n=6)})), (\ref{modbaryon:m>Gamma}) and
(\ref{modbaryon:Hos>Gamma}) determine which of the following scenario is
realized.

In the case (a), early oscillation begins in radiation dominated regime
and the baryon-to-moduli ratio (\ref{modbaryon:n_B/rho_chi(i)}) is
written as
\begin{equation}
  \frac{n_B}{\rho_\chi}=
   \frac{\delta_e m_{3/2} |\phi_{\mathrm{os}}|^2}{m_\chi^2 \chi_0^2}
   \left ( \frac{m_\chi}{H_{\mathrm{os}}} \right )^{3/2},
\end{equation}
where $\delta_e (\sim O(1))$ denotes the effective CP phase.  Note that as
far as the initial amplitude of the Affleck-Dine field is smaller than
$M_P$, it never dominates the universe at the instant of oscillation.
We can estimate $\phi_{\mathrm{os}}$ and $H_{\mathrm{os}}$ as
\begin{equation}
   |\phi|_{\mathrm{os}} \sim \gamma_*^{-\frac{1}{2}}\alpha M_P
\end{equation}
and
\begin{equation}
\begin{split}
   H_{\mathrm{os}} \sim \frac{\alpha^2 M_P^2}{\gamma_* M} 
     ~~~~~\mathrm{for}~~~n=4,\\
   H_{\mathrm{os}} \sim \frac{\alpha^4 M_P^4}{\gamma_*^2 M^3} 
     ~~~~~\mathrm{for}~~~n=6.
\end{split}
\end{equation}
where $\gamma_*=(\pi^2g_*(T_R)/90) \sim 25$.  Substituting these values
and using Eq.~(\ref{modbaryon:T_chi}), we finally obtain the
baryon-to-entropy ratio after decay of the modulus field as
\begin{equation}
\begin{split}
	\frac{n_B}{s}&=\frac{0.2\delta_e \sqrt{c\gamma_*}}{\alpha}
	\frac{m_{3/2}m_\chi}{M_P^2}\left( \frac{M}{M_P}\right)^{3/2}
	\left ( \frac{M_P}{\chi_0} \right )^2\\
	&\sim 7 \times 10^{-27}\delta_e \sqrt{c}
	\left( \frac{0.1}{\alpha}\right) 
	\left( \frac{m_{3/2}}{100 \mathrm{TeV}}\right)
	\left( \frac{m_{\chi}}{100 \mathrm{TeV}}\right)
	\left( \frac{M}{M_P}\right)^{3/2}  
	\left ( \frac{M_P}{\chi_0} \right )^2
\end{split}
\end{equation}
in the case of $n=4$ flat direction.  Clearly this is too small and it
is impossible that we obtain a proper amount of baryon asymmetry.  For
$n=6$, we obtain
\begin{equation}
\begin{split}
	\frac{n_B}{s}&=\frac{0.2\delta_e \sqrt{c}\gamma_*^2 }{\alpha^4}
	\frac{m_{3/2}m_\chi}{M_P^2}\left( \frac{M}{M_P}\right)^{9/2}
	\left ( \frac{M_P}{\chi_0} \right )^2\\
	&\sim 2 \times 10^{-21}\sqrt c \delta_e
	\left( \frac{0.1}{\alpha}\right) ^4
	\left( \frac{m_{3/2}}{100 \mathrm{TeV}}\right)
	\left( \frac{m_{\chi}}{100 \mathrm{TeV}}\right)
	\left( \frac{M}{M_P}\right)^{9/2}
	\left ( \frac{M_P}{\chi_0} \right )^2  .
\end{split}
\end{equation}
It also seems too small, but dependence of the cut-off scale $M$ is
rather large, so that if we assume $M\sim 100M_P$ the desired amount of
baryon asymmetry can be obtained.  It may seem peculiar that the cut-off
scale $M$ is bigger than Planck-scale, but our definition of $M$
includes some coupling constant, e.g., even if physical cut-off scale is
$M_P$, the effective cut-off scale can be $\sim 100M_P$ if the relevant
coupling constant is $10^{-2}$.

In the case (b), the modulus oscillation begins in radiation dominated era.
The baryon-to-moduli ratio (\ref{modbaryon:n_B/rho_chi(i)}) is expressed as
\begin{equation}
   \frac{n_B}{\rho_\chi}
     =\frac{\delta_e m_{3/2} |\phi_{\mathrm{os}}|^2}{m_\chi^2\chi_0^2}
	\left ( \frac{\Gamma_I}{H_{\mathrm{os}}} \right )^2
	\left ( \frac{m_\chi}{\Gamma_I} \right )^{3/2}.
\end{equation}
A straightforward calculation yields
\begin{equation}
\begin{split}
	\frac{n_B}{s}&=\frac{0.5\delta_e \sqrt{c}}{\alpha}
	\frac{m_{3/2}m_\chi}{M_P^2}\left( \frac{M}{M_P}\right)^{3/2}
	\left ( \frac{M_P}{\chi_0} \right )^2\\
	&\sim 8\times 10^{-27}\delta_e \sqrt{c}
	\left( \frac{0.1}{\alpha}\right) 
	\left( \frac{m_{3/2}}{100 \mathrm{TeV}}\right)
	\left( \frac{m_{\chi}}{100 \mathrm{TeV}}\right)
	\left( \frac{M}{M_P}\right)^{3/2}  
	\left ( \frac{M_P}{\chi_0} \right )^2
\end{split}
\end{equation}
for $n=4$ case. 
Obviously, this is too small.
On the other hand, for $n=6$ case we obtain
\begin{equation}
\begin{split}
	\frac{n_B}{s}&=\frac{0.5\delta_e \sqrt{c}}{\alpha^2}
	\frac{m_{3/2}m_\chi}{T_R M_P}\left( \frac{M}{M_P}\right)^{3}
	\left ( \frac{M_P}{\chi_0} \right )^2\\
	&\sim 2\times 10^{-16}\delta_e \sqrt{c}
	\left( \frac{0.1}{\alpha}\right)^2 
	\left( \frac{m_{3/2}}{100 \mathrm{TeV}}\right)
	\left( \frac{m_{\chi}}{100 \mathrm{TeV}}\right)
	\left( \frac{10^9\mathrm{GeV}}{T_R} \right)
	\left( \frac{M}{M_P}\right)^{3} 
	\left ( \frac{M_P}{\chi_0} \right )^2 .
\end{split}
\end{equation}
It seems possible that a proper amount of baryon asymmetry after
choosing cut-off scale appropriately.  But $T_R$ is constrained from the
condition $H_{\mathrm{os}}>m_\chi$ [Eq.(\ref{modbaryon:Hos>m(n=6)})].
Substituting Eq.~(\ref{modbaryon:Hos>m(n=6)}) into the above equation,
we obtain the upper bound on $n_B/s$,
\begin{equation}
\begin{split}
	\frac{n_B}{s} &\lesssim 
	\frac{0.5\delta_e \sqrt{c}}{\alpha}
	\frac{m_{3/2}^{1/4}m_\chi}{M_P^{5/4}}
	\left( \frac{M}{M_P}\right)^{9/4}
	\left ( \frac{M_P}{\chi_0} \right )^2\\
	&\sim 
	8\times 10^{-17}\delta_e \sqrt{c}
	\left( \frac{0.1}{\alpha}\right)^2 
	\left( \frac{m_{3/2}}{100 \mathrm{TeV}}\right)^{1/4}
	\left( \frac{m_{\chi}}{100 \mathrm{TeV}}\right)
	\left( \frac{M}{M_P}\right)^{9/4} 
	\left ( \frac{M_P}{\chi_0} \right )^2
\end{split}
\end{equation}
Thus we need $M\gtrsim 200M_P$ to obtain enough baryon asymmetry.
If this is the case, $T_R$ must also be as high as $10^{12}$ GeV.
This also satisfies the constraint $\Gamma_I>m_\chi$.

In the case (c), the modulus starts to oscillate in inflaton-dominated
regime and then inflaton decays resulting in brief radiation dominated
era followed by moduli dominated universe.  The baryon-to-moduli ratio
(\ref{modbaryon:n_B/rho_chi(i)}) in this case is expressed as
\begin{equation}
   \frac{n_B}{\rho_\chi}
    =\frac{\delta_e m_{3/2} |\phi_{\mathrm{os}}|^2}{m_\chi^2\chi_0^2}
	\left ( \frac{m_\chi}{H_{\mathrm{os}}} \right )^2.
\end{equation}
For the $n=4$ case, we obtain
\begin{equation}
\begin{split}
   \frac{n_B}{s}&=\frac{0.2\delta_e \sqrt{c}}{\alpha}
    \frac{m_{3/2}m_\chi^{3/2}}{T_R M_P^{3/2}}
        \left( \frac{M}{M_P}\right)^{3/2}
        \left ( \frac{M_P}{\chi_0} \right )^2\\
	&\sim 2\times 10^{-24}\delta_e \sqrt{c}
	\left( \frac{0.1}{\alpha}\right) 
	\left( \frac{m_{3/2}}{100\mathrm{TeV}}\right)
	\left( \frac{m_{\chi}}{100\mathrm{TeV}}\right)^{3/2}
	\left( \frac{10^9\mathrm{GeV}}{T_R} \right)
	\left( \frac{M}{M_P}\right)^{3/2}
	\left ( \frac{M_P}{\chi_0} \right )^2  ,
\end{split}
\end{equation}
which is extremely small compared with the present baryon density.
When we apply to the $n=6$ flat direction, the baryon-to-entropy ratio
is estimated as
\begin{equation}
\begin{split}
   \frac{n_B}{s}&=\frac{0.2\delta_e \sqrt{c}}{\alpha ^2}
    \frac{m_{3/2}m_\chi^{3/2}}{T_R^2 M_P^{1/2}}
       \left( \frac{M}{M_P}\right)^{3}
       \left ( \frac{M_P}{\chi_0} \right )^2\\
	&\sim 4\times 10^{-14}\delta_e \sqrt{c}
	\left( \frac{0.1}{\alpha}\right)^2 
	\left( \frac{m_{3/2}}{100 \mathrm{TeV}}\right)
	\left( \frac{m_{\chi}}{100 \mathrm{TeV}}\right)^{3/2}
	\left( \frac{10^9 \mathrm{GeV}}{T_R} \right)^2
	\left( \frac{M}{M_P}\right)^{3}
	\left ( \frac{M_P}{\chi_0} \right )^2 ,
\end{split}
\end{equation}
which seems successful. However, we need rather high reheating
temperature which suppress the baryon-to-entropy ratio, due to the
condition $H_{\mathrm{os}}>m_\chi$ [Eq.(\ref{modbaryon:Hos>m(n=6)})].
Substituting Eq.(\ref{modbaryon:Hos>m(n=6)}), the upper limit for baryon
asymmetry is obtained,
\begin{equation}
\begin{split}
	\frac{n_B}{s} &\lesssim 0.2\delta_e \sqrt{c}
	\frac{m_{3/2}}{M_P}
	\left( \frac{M}{M_P}\right)^{3/2}\\
	&\sim 
	8\times 10^{-15}\delta_e \sqrt{c}
	\left( \frac{0.1}{\alpha}\right)^2 
	\left( \frac{m_{3/2}}{100\mathrm{TeV}}\right)
	\left( \frac{M}{M_P}\right)^{3/2} .
\end{split}
\end{equation}
If $M\sim 100M_P$ we can obtain desired baryon asymmetry, and this
indicates that reheating temperature should be higher than $\sim
10^{11}$GeV.  On the other hand, $T_R\lesssim 10^{12}$GeV is necessary
in order to satisfy $\Gamma_I <m_\chi$.  Thus $M\sim
100M_P$ and $10^{11}$GeV $\lesssim T_R \lesssim 10^{12}$GeV are the
possible parameter region (see Fig.~\ref{fig:m_chi=1e5}).

\subsubsection{$H_{\mathrm{os}}<m_\chi$}

Now let us turn to the case $H_{\mathrm{os}}<m_\chi$.  As we explained,
early oscillation to avoid charge or color breaking minima requires
$H_{\mathrm{os}}>m_{3/2}$ and hence this particular possibility arises
only when modulus mass $m_\chi$ is much heavier than $m_{3/2}$
\footnote{As we explain in Sec.~\ref{sec:remarks}, although moduli decay
into gravitinos may cause cosmological difficulty, here gravitinos are
also heavy enough to decay well before the BBN.  Furthermore LSPs
produced by decay of moduli effectively annihilates and do not overclose
the universe (or they become dark matter).
However, the subsequent decay of non-thermally produced gravitinos
may pose a cosmological difficulty. See Sec.~\ref{sec:remarks}.}.  
In this case, we can
classify the cosmological scenario depending on the inflaton decay rate
$\Gamma_I$ : (d) $\Gamma_I>m_\chi>H_{\mathrm{os}}$, (e)
$m_\chi>\Gamma_I>H_{\mathrm{os}}$, (f)
$m_\chi>H_{\mathrm{os}}>\Gamma_I$.  Note that baryon-to-moduli ratio is
fixed once the Affleck-Dine field starts to oscillate, but resulting
formula for $n_B/\rho_\chi$ is the same as
eq.(\ref{modbaryon:n_B/rho_chi(i)}).  Therefore the results of case (d)
and (f) are the same as(a) and (c) respectively.  Only the case (e)
slightly differs from (b).

In the case (e) the baryon-to-moduli ratio is expressed as
\begin{equation}
     \frac{n_B}{\rho_\chi}=
       \frac{\delta_e m_{3/2} |\phi_{\mathrm{os}}|^2}{m_\chi^2\chi_0^2}
	\left ( \frac{m_\chi}{\Gamma_I} \right )^2
	\left ( \frac{\Gamma_I}{H_{\mathrm{os}}} \right )^{2},
\end{equation}
which is slightly different from the case (b).
Note that we have used the approximation that the moduli dominate the universe
soon after the oscillation.
The following calculations are similar, and the result is
\begin{equation}
\begin{split}
       \frac{n_B}{s}&=\frac{0.2\delta_e \sqrt{c}\gamma_*^4}{\alpha ^{6}}
	\frac{m_{3/2}m_\chi^{3/2}T_R^4}{ M_P^{13/2}}
	\left( \frac{M}{M_P}\right)^{4}
	\left ( \frac{M_P}{\chi_0} \right )^2 \\
	&\gtrsim 4 \times 10^{-30}\delta_e \sqrt{c}
	\left( \frac{0.1}{\alpha}\right)^2 
	\left( \frac{m_{3/2}}{100 \mathrm{TeV}}\right)
	\left( \frac{m_{\chi}}{100 \mathrm{TeV}}\right)^{3/2}
	\left( \frac{M}{M_P}\right)^{2}  
	\left ( \frac{M_P}{\chi_0} \right )^2
\end{split}
\end{equation}
for $n=4$ case, where we have used the constraint $\Gamma_I > H_{\rm os}$ in the second line. 
Using the same constraint, for $n=6$ case we obtain
\begin{equation}
\begin{split}
	\frac{n_B}{s}&=\frac{0.2\delta_e \sqrt{c}\alpha^{18}}{\gamma_*^{12}}
	\frac{m_{3/2}m_\chi^{3/2} M_P^{19/2}}{T_R^{12}}
	\left( \frac{M_P}{M}\right)^{12}
	\left ( \frac{M_P}{\chi_0} \right )^2\\
	&\lesssim 4\times 10^{-33}\delta_e \sqrt{c}
	\left( \frac{0.1}{\alpha}\right)^{6} 
	\left( \frac{m_{3/2}}{100 \mathrm{TeV}}\right)
	\left( \frac{m_{\chi}}{100 \mathrm{TeV}}\right)^{3/2}
	\left( \frac{M}{M_P}\right)^{6}
	\left ( \frac{M_P}{\chi_0} \right )^2 .
\end{split}
\end{equation}
Similar to the case for $H_{\rm os} > m_{\chi}$, for appropriate choice
of the cut-off scale $M$ and the reheating temperature $T_R$, it seems
that we can obtain a proper amount of baryon asymmetry.  However, we
should recall that the constraint $m_{3/2} < H_{\mathrm{os}} <m_\chi$
narrows the allowed parameter range.
In fact, the case (e) is not realized in the parameter region we are interested in.

In Figs.\ref{fig:m_chi=1e5} and \ref{fig:m_chi=1e7} we show the
resulting baryon-to-entropy ratio in $(M,T_R)$ plane in the case of
$m_{3/2}=m_{\chi}=100~$TeV and $m_\chi=(4\pi^2)
m_{3/2}=(4\pi^2)100~$TeV for $n=6$.  The latter case is naturally realized in
mirage-mediation models.  We can see that $T_R \gtrsim 10^{11}~$GeV and
$M\gtrsim 10^{20}~$GeV are required in the former case.  In the latter
case where the modulus field is much heavier than the gravitino, the
constraint is weaker.  Note that in such a heavy moduli scenario
gravitinos can be efficiently produced by the decay of moduli, and these
non-thermal gravitinos also decay before BBN for $m_{3/2}\sim 100~$TeV.
LSPs produced by the decay of those gravitinos may be harmful.  We will
discuss it in Sec.~\ref{sec:remarks}.

\subsection{Q-ball formation }

Finally we must consider the effects of Q-ball formation.  The
fluctuations of the Affleck-Dine field with $U(1)_B$ charge grow and
result in lumped condensate, called Q-balls
\cite{Coleman:1985ki,Kusenko:1997zq}.  The Q-ball formation leads to
many non-trivial cosmological consequences, and they highly depend on
SUSY breaking models \cite{Kusenko:1997si,Enqvist:1997si} (see also
\cite{Kasuya:2001hg,Fujii:2002kr}).  As we have seen in the previous
subsection, quite large cut-off scale $M$ is required.  One may wonder
this leads to large Q-balls and invalidates the applicability of our
scenario.  However, as we will see, largeness of Q-balls is suppressed
because of early oscillation.  The radius of Q-balls is comparable to
the hubble horizon scale at the epoch of Q-ball formation.  Thus
although larger cut-off scale $M$ tends to create larger Q-balls, but
higher reheating temperature $T_R$, which causes earlier oscillation,
tends to make Q-balls smaller.  Now let us estimate $Q$.

It is found that that for the Q-balls which have developed via
logarithmic potential, the total charge of Q-ball $Q$ is fitted by the
formula~\cite{Kasuya:2001hg},
\begin{equation}
   Q=\beta \left ( 
	    \frac{|\phi_{\mathrm{os}}|}{T_{\mathrm{os}}} \right )^4
\end{equation}
where $\beta \sim 6\times 10^{-4}$.
Applying to the early oscillation case for $n=6$ flat direction, 
it is estimated as 
\begin{equation}
	Q\sim 4 \times 10^{17}
	\left ( \frac{\beta}{6\times 10^{-4}} \right )
	\left( \frac{10^{11} \mathrm{GeV}}{T_R} \right)^{2}
	\left( \frac{M}{100M_P}\right)^{3}
\end{equation}
for $H_{\rm os} > \Gamma_I$, and
\begin{equation}
	Q\sim 9 \times 10^{16}
	\left ( \frac{\beta}{6\times 10^{-4}} \right )
	\left( \frac{0.1}{\alpha} \right)^{4}
	\left( \frac{M}{100M_P}\right)^{6}
\end{equation}
for $H_{\rm os} < \Gamma_I$.
It is known that evaporation of Q-balls in high-temperature plasma can
efficiently transfer the charge of Q-balls up to $\Delta Q \sim 10^{20}$
almost model independently \cite{Laine:1998rg} (see also
\cite{Kawasaki:2006yb}).  Therefore in the most interesting parameter
region, Q-balls formed through Affleck-Dine mechanism can completely
evaporate and have no further effects on cosmological evolution of
baryon asymmetry.

\begin{figure}[htbp]
   \begin{center}
    \includegraphics[width=0.65\linewidth]{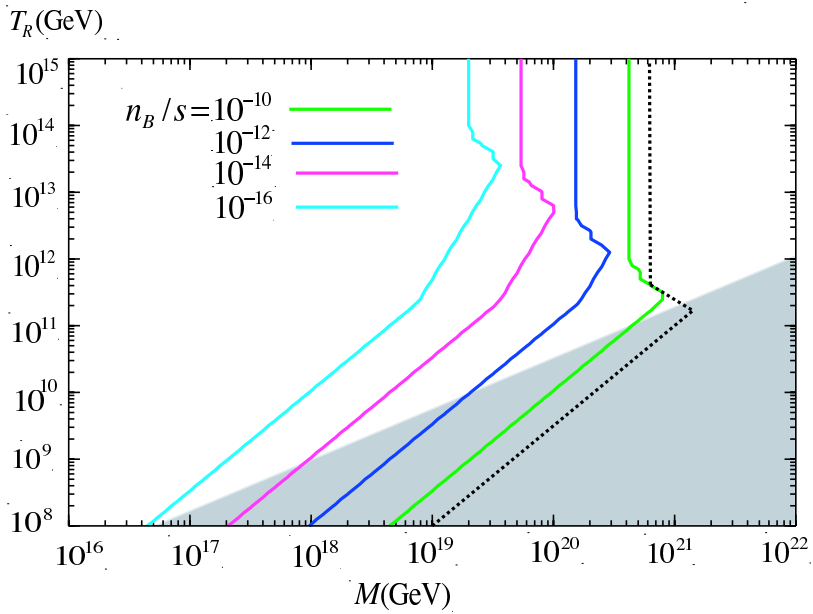}
    \caption{Contour plot of $n_B/s$ as a function of $M$ and $T_R$.  We
    take $m_\chi=m_{3/2}=100$TeV.  In the shaded region Affleck-Dine
    field is trapped into charge-breaking minima
    and baryogenesis does not work.
    Also we show by the dotted line $Q\sim 10^{20}$. 
    The left side of the dotted line predicts $Q<10^{20}$ and 
    Q-balls completely evaporate in high-temperature plasma.}
    \label{fig:m_chi=1e5}
   \end{center}
\end{figure}

\begin{figure}[htbp]
 \begin{center}
  \includegraphics[width=0.65\linewidth]{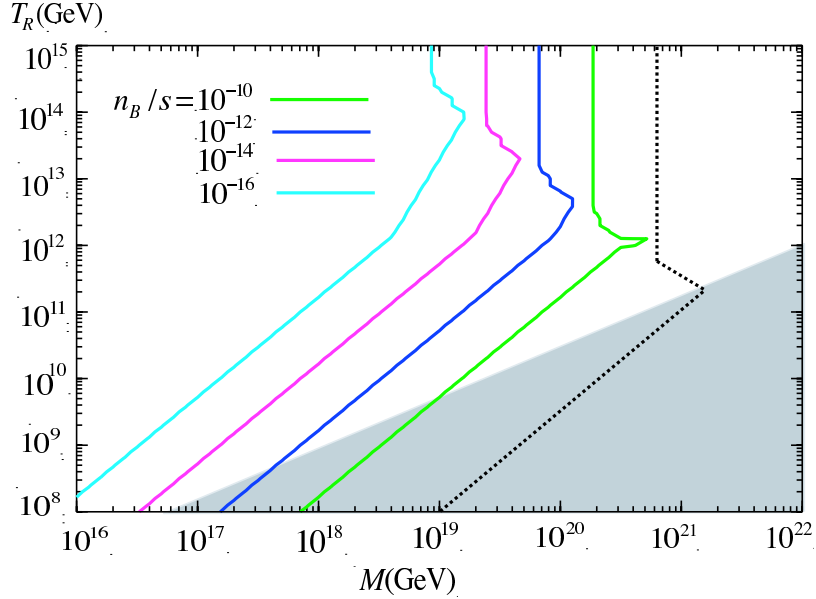}
   \caption{ Same as Fig.\ref{fig:m_chi=1e5}, except for 
  $m_\chi =(4\pi^2)m_{3/2} =(4\pi^2)100$TeV.  }
  \label{fig:m_chi=1e7}
 \end{center}
\end{figure}

In Figs.\ref{fig:m_chi=1e5} and \ref{fig:m_chi=1e7}, we show the contour
of $Q\sim 10^{20}$ with black dotted line.  It can be seen that in the
interesting parameter region where $n_B/s\sim 10^{-10}$ is obtained,
only small Q-balls are produced and they evaporate in the
high-temperature plasma.


\section{Affleck-Dine baryogenesis with gauged $U(1)_{B-L}$}  
\label{sec:gauged}

Next we turn to another possibility that Affleck-Dine baryogenesis with
large gravitino mass works with an extension of MSSM to include some
additional fields and gauged $U(1)_{B-L}$ symmetry.  Because the global
$U(1)_{B-L}$ symmetry within MSSM is anomaly-free, it can naturally be
extended to local symmetry.  But from the viewpoint of baryogenesis, it
must be spontaneously broken at some high energy scale in order to
create baryon asymmetry and not to contradict with terrestrial
experiments such as proton decay.

\subsection{The model}

We briefly explain the model discussed in Ref.~\cite{Fujii:2001sn}.
First, we introduce MSSM singlet fields which have the superpotential as
\begin{equation}
	W = \lambda X(S \bar S -v^2),
\end{equation}
where $X, S$ and $\bar S$ have the $U(1)_{B-L}$ charge $0,2$ and $-2$
respectively, and $v$ denotes the $U(1)_{B-L}$ breaking scale.  They
induce the scalar potential given by
\begin{equation}
\begin{split}
  V =& |\lambda |^2 
    \left \{ |X|^2 (|S|^2 +|\bar S|^2 )+ |S\bar S- v^2|^2 \right \} \\
  &+ \frac{g^2}{2}\left ( 2|S|^2 -2|\bar S|^2 - q |\phi|^2 \right )^2
\end{split}
\end{equation}
where $g$ denotes the $U(1)_{B-L}$ gauge coupling constant and $q$
denotes the $U(1)_{B-L}$ charge of the Affleck-Dine field.  The second
term comes from the $D$-term contribution.  In the following, we
consider flat directions which are lifted by $n=6$ non-renormalizable
superpotential in MSSM, such as $udd$ or $LLe$ direction.  In this
model, gauge-invariant superpotential which lifts those flat directions
are given by
\begin{equation}
	\frac{k_1}{6M^3}\left ( \frac{S}{M} \right )(udd)^2,~~~~~
	\frac{k_2}{6M^3}\left ( \frac{S}{M} \right )(LLe)^2
\end{equation}
where $k_1$ and $k_2$ are $O(1)$ coupling constants,
and the resulting zero-temperature scalar potential is written as
\begin{equation}
\begin{split}
  V=& m_\phi^2 |\phi|^2 -c_{\rm H}H^2|\phi|^2
    +\frac{m_{3/2}}{6M^3}\left ( \frac{S}{M} \right )
      (a_m\phi^6+\mathrm{h.c.})
    +\frac{H}{6M^3}\left ( \frac{S}{M} \right )
      (a_H\phi^6+\mathrm{h.c.})\\
    &+\frac{1}{M^6}\left ( \frac{S}{M} \right )^2|\phi|^{10}
     +\frac{1}{36M^8}|\phi|^{12}.
\end{split}
\end{equation}
Although the whole dynamics is somewhat complicated and we do not give
the details here (see \cite{Fujii:2001sn} for a detail), the point is
that by using the additional $D$-term potential which does not exist in
MSSM, the Affleck-Dine field can be stopped at the $U(1)_{B-L}$ breaking
scale $v$ during inflation.  If $v$ is smaller than the hill of the
potential of the Affleck-Dine field
\begin{equation}
   v \lesssim |\phi|_{\mathrm{hill}} \sim 
      \left ( \frac{m_\phi^2M^{4}}{m_{3/2}\langle S \rangle} 
	 \right )^{1/4},
\end{equation}
Affleck-Dine mechanism works without trapping into the charge or color
breaking global minimum.  If we assume $\langle S \rangle \sim v$ and we
focus on $n=6$ case, this condition is equivalent to
\begin{equation}
 \begin{split}
  v &\lesssim \left ( \frac{m_\phi^2 M^4}{m_{3/2}} \right )^{1/5} \\
  &\sim 8\times 10^{14}\mathrm{GeV}
  \left ( \frac{100 \mathrm{TeV}}{m_{3/2}} \right )^{1/5}
  \left ( \frac{m_\phi}{1 \mathrm{TeV}} \right )^{2/5}
  \left ( \frac{M}{M_P} \right )^{4/5} .   \label{modbaryon:vbound}
 \end{split}
\end{equation}
If the value $v$ exceeds this bound, Affleck-Dine baryogenesis can not
work due to trapping of the Affleck-Dine field in global charge breaking
minima, if thermal effects are neglected.

\subsection{Baryon asymmetry}

We saw that in this type of model, the Affleck-Dine field stops at
$U(1)_{B-L}$ breaking scale $v$ until Hubble parameter becomes of the
order $m_\phi$ and oscillation begins.  If $v$ is smaller than the hill
of the potential of the Affleck-Dine field, Affleck-Dine mechanism
works.  In the case of early oscillation, the result is the same as
usual early oscillation scenario considered in the previous section.
Thus we consider only the case of no early oscillation in this
subsection.  The condition to avoid early oscillation is
\begin{equation}
\begin{split}
	T_R & \lesssim \frac{m_\phi^{1/2}v}{\alpha M_P^{1/2}} \\
	& \sim 2.1\times 10^8\mathrm{GeV}
	\left ( \frac{0.1}{\alpha} \right )
	\left ( \frac{m_\phi}{1 \mathrm{TeV}} \right )^{2/5}
	\left ( \frac{v}{10^{15} \mathrm{GeV}} \right ).
\end{split}
\end{equation}
Thus we can safely set $\Gamma_I < m_\chi$.  The baryon number
at the instant of oscillation of the Affleck-Dine field is given by
\begin{equation}
  n_{B}(t_{\mathrm{os}})=\frac{4\beta |a_m|}{9}
  \frac{\delta_e m_{3/2}}{H_{\mathrm{os}}M^4}v^7
\end{equation}
with $H_{\mathrm{os}}\sim m_\phi$.  The baryon-to-moduli ratio is once
fixed at the epoch of oscillation of the Affleck-Dine field,
$t=t_{\mathrm{os}}$, and the final reheating comes from the decay of
moduli.  The result is
\begin{equation}
   \frac{n_B}{s}=0.1\sqrt c \delta_e
   \frac{m_{3/2}m_\chi^{3/2}v^7}{m_\phi^3M^4M_P^{5/2}}
   \left ( \frac{M_P}{\chi_0} \right )^2,
\end{equation}
which depends on seventh powers of $v$.  Substituting the upper bound on
$v$ [Eq.(\ref{modbaryon:vbound})], we obtain an upper bound on the
baryon-to-entropy ratio,
\begin{equation}
 \begin{split}
  \frac{n_B}{s} & \lesssim
  0.1\sqrt c \delta_e 
  \frac{m_\chi^{3/2}M^{8/5}}{m_\phi^{1/5}m_{3/2}^{2/5}M_P^{5/2}}
  \left ( \frac{M_P}{\chi_0} \right )^2\\
  & \sim 5\times 10^{-13}\sqrt c \delta_e
  \left ( \frac{m_\chi}{100 \mathrm{TeV}} \right )^{3/2}
  \left ( \frac{100 \mathrm{TeV}}{m_{3/2}} \right )^{2/5}
  \left ( \frac{1 \mathrm{TeV}}{m_\phi} \right )^{1/5}
  \left ( \frac{M}{M_P} \right )^{8/5}
  \left ( \frac{M_P}{\chi_0} \right )^2,
 \end{split}
\end{equation}
which seems successful.  However, it is non-trivial whether Q-ball is
small enough to evaporate completely.  Charge of Q-ball is given by \cite{Kasuya:2000wx}
\begin{equation}
  Q\sim \gamma \left ( \frac{v}{m_\phi} \right )^2 \times \left \{
	     \begin{array}{ll}
	      \epsilon~~~~~&(\epsilon \gtrsim 0.01 )\\
		0.01  ~~~~~&(\epsilon \lesssim 0.01)
	     \end{array}
	\right.    \label{Qcharge:grav}
\end{equation}
where $\gamma$ is order
$10^{-2}-10^{-3}$ factor which represents the delay of Q-ball formation from the
oscillation of Affleck-Dine field and $\epsilon$ is called the ellipticity parameter given by
\begin{equation}
	\epsilon \sim \delta_e \frac{m_{3/2}v^5}{m_\phi^2 M^4}.
\end{equation}
Therefore,
using the upper bound of $v$ [Eq.(\ref{modbaryon:vbound})], we obtain
\begin{equation}
 \begin{split}
  Q & \sim
  \frac{4 \delta_e \gamma}{9}\frac{m_{3/2}v^7}{m_\phi^4 M^4}\\
  & \lesssim 1\times 10^{21} \delta_e 
  \left ( \frac{\gamma}{6\times 10^{-3}} \right )
  \left ( \frac{100\mathrm{TeV}}{m_{3/2}} \right )^{2/5}
  \left ( \frac{1\mathrm{TeV}}{m_\phi} \right )^{6/5}
  \left ( \frac{M}{M_P} \right )^{8/5},
 \end{split}
\end{equation}
for $\epsilon \gtrsim 0.01$,
which is a little larger than total evaporated charge $\Delta Q \sim
10^{20}$.  For pure leptonic flat direction such as $LLe$, Q-balls must
completely evaporate above the temperature where electroweak phase
transition occurs in order to convert lepton number into baryon number
by sphaleron effects, and hence $Q \gtrsim 10^{20}$ is not acceptable.
On the other hand for flat directions carrying baryon number such as
$udd$, $10^{20} \lesssim Q\lesssim 10^{22}$ is allowed.  In such a case,
where Q-balls decay below the freeze-out temperature of LSP, we must
care about overproduction of LSPs from Q-ball decay.  But in our
scenario entropy production from moduli decay dilutes them.  Therefore
for $udd$ direction \footnote{Actually $LLe$ and $udd$ direction can
have large field value simultaneously.}  gauged U(1)$_{B-L}$ scenario in
the presence of heavy moduli is marginally possible.

\begin{figure}[htbp]
 \begin{center}
  \includegraphics[width=0.65\linewidth]{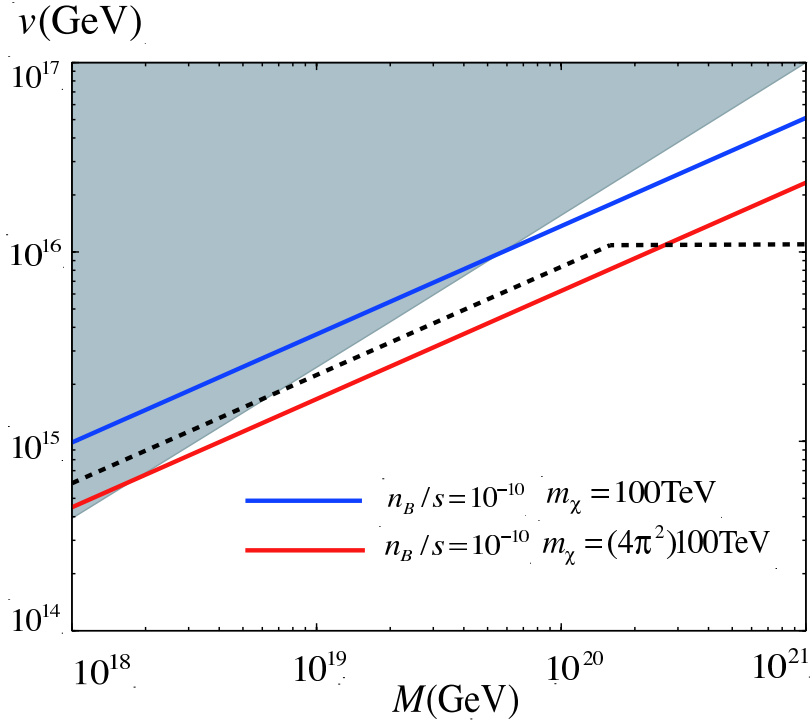} \caption{The two
  solid lines show $n_B/s \sim 10^{-10}$ in gauged $U(1)_{B-L}$
  scenario, the upper (blue) line corresponds to $m_\chi = 100$ TeV and
  the lower (red) line corresponds to $m_\chi = (4\pi^2)100$ TeV.  We
  take $m_{3/2}=100$ TeV.  In the dark shaded region Affleck-Dine field
  is trapped into charge-breaking minima and baryogenesis does not work.
  We also show $Q\sim 10^{22}$ by the dotted line.}
  \label{fig:gaugedAD}
 \end{center}
\end{figure}

In Fig.~\ref{fig:gaugedAD}, we show the resultant baryon asymmetry in
($M,v$) plane with constraints.  We can see that for $m_\chi =100$
TeV, Q-balls become too large.  But for larger $m_\chi$ the correct
baryon asymmetry can be obtained without forming too large Q-balls.


\section{Affleck-Dine baryogenesis without superpotential}  
\label{sec:W=0}

\subsection{The model}

Next we consider the models of Affleck-Dine baryogenesis with gauged
$U(1)_{B-L}$ including no non-renormalizable superpotentials due to some
symmetry such as $R$-symmetry.  In such a case, baryon number violating
operators are supplied by higher order effects from Kahler potentials
(see e.g., Ref.~\cite{Fujii:2002aj}) and the initial amplitude of the
Affleck-Dine field can become as large as Planck scale.
The dynamics of the Affleck-Dine field is similar to the previous
section.  As a result, large Q-balls are formed associated with
Affleck-Dine baryogenesis and they decay at late time after the
freeze-out of LSPs but before BBN.  Interestingly, in this type of model
late-decaying Q-balls may once dominate the universe
\cite{Fujii:2002aj}.  If this is the case, a nice feature arises when
considering the moduli-induced gravitino problem, as explained in
Sec.~\ref{sec:remarks}.

Now let us investigate the above model.  The zero-temperature scalar
potential for the flat direction $\psi$ is given by
\begin{equation}
 \begin{split}
  V(\psi) = &(m_\psi^2-c_H H^2)|\psi|^2+
  \frac{m_{3/2}^2}{nM^{n-2}}(a_m \psi^n + \mathrm{h.c.}) \\
  &+ \frac{H^2}{nM^{n-2}}(a_H \psi^n + \mathrm{h.c.}) + \dots
  \label{pot:W=0}
 \end{split}
\end{equation}
where the ellipsis denote the higher order terms, which stabilize the
Affleck-Dine field at some value of order the Planck-scale.  Note that
the potential (\ref{pot:W=0}) also has charge and/or color breaking
global minimum near the field value at $M$.  Similar to the previous
section, in order to avoid falling into this minimum, the $D$-term
stopping at $v$ must satisfy the following condition,
\begin{equation}
   v \lesssim |\psi|_{\mathrm{hill}}\sim \frac{m_\psi}{m_{3/2}}M.
\end{equation}
Here we consider only the case without early oscillation due to thermal
effects.  This requires
\begin{equation}
 \begin{split}
  T_R & \lesssim \alpha ^{-1} m_\psi ^{1/2}|\psi_0| M_P^{-1/2} \\
  & \sim 2 \times 10^{11} \mathrm{GeV} 
  \left ( \frac{0.1}{\alpha } \right )
  \left( \frac{m_{\psi}}{100 \mathrm{GeV}} \right)^{1/2}
  \left( \frac{|\psi_0|}{M_P}\right ).
 \end{split} 
\end{equation}
At the beginning of the oscillation $H = m_\psi$,
the baryon number is calculated as
\begin{equation}
   n_B(t_{\mathrm{os}}) \simeq 
    \frac{|a_m|\delta_e m_{3/2}^2}{m_\psi M^{n-2}}\psi_0^n
\end{equation}
where 
$\psi_0$ is given by $U(1)_{B-L} $ breaking scale $v$.  Note that we
assume the Affleck-Dine field has baryonic charge.  In such a case, the
whole baryon number created by the coherent motion of the Affleck-Dine
field contributes to the baryon number of the universe as far as Q-balls
decay before BBN.  If it does not have baryonic charge and only has
leptonic charge, only some fraction of the total created lepton number
evaporated from Q-balls at the temperature above the electroweak scale
can be converted into baryon number through the sphaleron effects
\cite{Kuzmin:1985mm,Khlebnikov:1988sr}.  Thus the resultant baryon
asymmetry is suppressed.  We do not consider such a case.

The charge of Q-balls is given by Eq.~(\ref{Qcharge:grav}) 
where the ellipticity parameter $\epsilon$ is now estimated as
\begin{equation}
	\epsilon \sim \delta_e \left ( \frac{m_{3/2}}{m_\psi} \right )^2 \left ( \frac{v}{M} \right )^{n-2}.
\end{equation}
Thus we obtain the charge of Q-balls
\begin{equation}
	Q \sim 4 \times 10^{26}
	\left ( \frac{\gamma}{6\times 10^{-3}} \right )
	\left( \frac{1 \mathrm{TeV}}{m_\psi} \right)^{2}
	\left ( \frac{v}{M_P} \right )^2.
\end{equation}
for $\epsilon \lesssim 0.01$, and
\begin{equation}
	Q \sim 4 \times 10^{32}
	\left ( \frac{\gamma}{6\times 10^{-3}} \right )
	\left ( \frac{m_{3/2}}{100 \mathrm{TeV} } \right )^2
	\left( \frac{1 \mathrm{TeV}}{m_\psi} \right)^{4}
	\left ( \frac{v}{M_P} \right )^n
	\left ( \frac{M_P}{M} \right )^{n-2}.
\end{equation}
for $\epsilon \gtrsim 0.01$.
On the other hand, the decay temperature of Q-balls is estimated as 
\cite{Cohen:1986ct,Kawasaki:2006yb}
\begin{equation}
	T_Q \sim 1.8\mathrm{GeV}
	\left(\frac{m_\psi}{1\mathrm{TeV}} \right)^{1/2}
	\left(\frac{10^{24}}{Q} \right)^{1/2}
	\label{T_Q}
\end{equation}
if there exist lighter scalar fields than the Affleck-Dine field,
and hence we can see that Q-balls decay below the electroweak scale but
before BBN for some parameter region, $v \ll M_P$ and/or $M\gg M_P$.
The entire cosmological scenario depends on the decay temperature of
Q-balls $T_Q$, the initial amplitude of the Affleck-Dine field $v$ and
that of the modulus filed $\chi_0$.  We assume the reheating temperature
from inflaton is not so high as inflaton dominates the universe
when the Affleck-Dine field begins to oscillate but decays well before
the modulus field decays.  This is satisfied for $10 \mathrm{GeV}
\lesssim T_R \lesssim 10^9 \mathrm{GeV}$.  The following analysis does
not depend on the precise value of $T_R$ as far as the $T_R$ lies in the
above range.

The final reheating comes from moduli or Q-balls.
If the following condition
\begin{equation}
	T_Q <  T_\chi \left ( \frac{v}{\chi_0} \right )^2
\end{equation}
is satisfied ($T_\chi$ is the decay temperature of the modulus field), 
Q-balls dominate the universe before they decay 
but after the modulus field decays.

\subsection{Baryon asymmetry}

In the case with Q-ball domination,
the baryon-to-entropy ratio is fixed at the decay of Q-balls,
\begin{equation}
	\frac{n_B}{s}=\frac{n_B}{\rho_\psi}\frac{\rho_\psi}{s}
	=\frac{n_B}{\rho_\psi}\frac{3T_Q}{4}
\end{equation}
and the subsequent cosmological scenario does not depend on the
properties of moduli.  This is calculated as
\begin{equation}
\begin{split}
	\frac{n_B}{s}\sim & 7 \times 10^{-3}
	\left ( \frac{6\times 10^{-3}}{\gamma} \right )^{1/2}
	\left( \frac{1 \mathrm{TeV}}{m_\psi} \right)^{3/2}\\
	& \times
	\left( \frac{m_{3/2}}{100 \mathrm{TeV}} \right)^2
	\left( \frac{M_P}{M} \right)^{2}
	\left ( \frac{v}{M_P} \right )
\end{split}
\end{equation}
for $\epsilon \lesssim 0.01$, and
\begin{equation}
\begin{split}
	\frac{n_B}{s}\sim & 2 \times 10^{-5}
	\left ( \frac{6\times 10^{-3}}{\gamma} \right )^{1/2}
	\left( \frac{1 \mathrm{TeV}}{m_\psi} \right)^{1/2}\\
	& \times
	\left( \frac{m_{3/2}}{100 \mathrm{TeV}} \right)
	\left( \frac{M_P}{M} \right)^{(n-2)/2}
	\left ( \frac{v}{M_P} \right )^{(n-4)/2},
\end{split}
\end{equation}
for $\epsilon \gtrsim 0.01$,
where we have assumed $\delta_e \sim 0.01$.
On the other hand, in the case of no Q-ball domination, 
the final reheating occurs due to the modulus decay.
The baryon-to-entropy ratio is thus given by
\begin{equation}
	\frac{n_B}{s}=\frac{n_B}{\rho_\chi}\frac{\rho_\chi}{s}
	=\frac{n_B}{\rho_\chi}\frac{3T_\chi}{4}
\end{equation}
and it is estimated as
\begin{equation}
\begin{split}
	\frac{n_B}{s}\sim & 2 \times 10^{-4}\sqrt{c}
	\left ( \frac{m_\chi}{100 \mathrm{TeV}} \right )^{3/2}
	\left( \frac{1 \mathrm{TeV}}{m_\psi} \right)^{3}\\
	& \times
	\left( \frac{m_{3/2}}{100 \mathrm{TeV}} \right)^2
	\left( \frac{M_P}{M} \right)^{n-2}
	\left ( \frac{v}{M_P} \right )^{n}
	\left ( \frac{M_P}{\chi_0} \right )^2.
\end{split}
\end{equation}
In both cases, we can see that 
$M \gg M_P$ and/or $v \ll M_P$ is required in order to obtain
correct order of baryon asymmetry.

\begin{figure}[thbp]
 \begin{center}
  \includegraphics[width=0.65\linewidth]{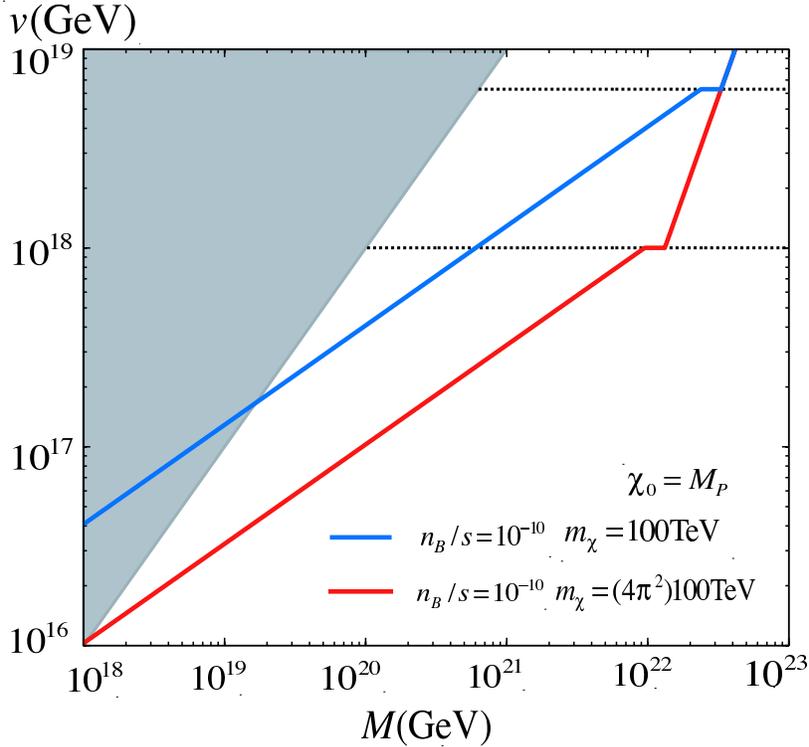} \caption{The two
  solid lines show $n_B/s \sim 10^{-10}$ for $m_\chi = 100$ TeV and
  $(4\pi^2)100$ TeV in the no superpotential model.  We take
  $m_{3/2}=100$ TeV.  In the dark shaded region Affleck-Dine field is
  trapped into charge-breaking minima and baryogenesis does not work.}
  \label{fig:W=0_1}
 \end{center}
\end{figure}

We show in Fig.~\ref{fig:W=0_1} the contour where the appropriate baryon
asymmetry $n_B/s \sim 10^{-10}$ is obtained for $n=4$ and the modulus
mass $m_\chi = 100$ TeV and $(4\pi^2)100$ TeV.  It should be noticed
that for $v > \sqrt 3 M_P, $ a brief period of inflation occurs due to
the Affleck-Dine field.  But in general, supergravity effects steepen
the potential above the Planck scale, and hence the region with $v
\gtrsim M_P$ is not favored from naturalness.  Above the dotted lines
Q-ball domination is realized for each modulus mass.   
It can be seen that for
$M\gtrsim 10^{18}$ GeV and $v \gtrsim 10^{16}$ GeV, this baryogenesis
mechanism works.  But another subtlety arises when one considers the LSP
produced by the Q-ball decay or gravitinos from modulus decay.  This
will be discussed in Sec.~\ref{sec:remarks}.


\section{Remarks on heavy modulus decay}  \label{sec:remarks}

Before closing the discussion, we consider some non-trivial feature of
the modulus decay.  In the above arguments, we have not considered the
details of the decay products of the modulus field.  We briefly discuss
the other consequences of modulus decay on cosmological evolution.

\subsection{Non-thermal dark matter from modulus decay}

One of the favored nature of the supersymmetric theory is that it can
provide the candidate of the dark matter of the universe.  Under the
R-parity conservation, the lightest supersymmetric particle (LSP)
becomes stable and if it has the appropriate annihilation cross section,
it can account for the energy density of the dark matter
\cite{Jungman:1995df}.  In anomaly-mediated SUSY breaking models,
wino-like neutralino naturally becomes the LSP.  In the standard thermal
relic scenario, the mass of wino should be as heavy as 2 TeV to account
for the dark matter because of its large annihilation cross section
\footnote{There is an argument that non-perturbative effect enhances the
annihilation cross section, and wino should be about 3 TeV if its thermal
relic accounts for the dark matter of the universe
\cite{Hisano:2006nn}.}.  In mirage-mediation model, the LSP is the mixed
state of bino and higgsino-like neutralino \cite{Endo:2005uy,Choi:2005uz} and their
thermal relic abundance can account for the dark matter of the universe
\cite{Baer:2006id}.  But in our scenario, the final reheating
temperature from modulus decay is typically much lower than the
freeze-out temperature of LSP and thermal relic can not be the dark
matter.  However, there is a non-thermal origin of the dark matter from
the decay of moduli, and there arises a possibility that non-thermal
LSPs can account for the dark matter of the universe.  Its abundance is
estimated as \cite{Moroi:1999zb,Fujii:2001xp}
\begin{equation}
   Y(T) \simeq \left [ \frac{1}{Y(T_\chi)}+ 
	\sqrt{\frac{8\pi^2g_*}{45}}\langle \sigma v \rangle 
	M_P(T_\chi-T) \right ]^{-1}
	\label{nonthermalLSP}
\end{equation}
where $Y=n_{\mathrm{LSP}}/s$ and $T_\chi$ denotes the decay temperature
of the modulus field.  We can see that if the annihilation cross section
is large enough, the second term dominates and the relic abundance is
inversely proportional to it.  In terms of the density parameter, we can
rewrite it as
\begin{equation}
   \Omega_{\mathrm{LSP}}h^2 \sim 0.27
    \left ( \frac{10}{g_*(T_\chi)} \right )^{1/2}
    \left ( \frac{100\mathrm{MeV}}{T_\chi} \right )
    \left ( \frac{m_{\mathrm{LSP}}}{100\mathrm{GeV}} \right )^3
    \left ( \frac{10^{-3}}{m_{\mathrm{LSP}}^2 \langle 
     \sigma v \rangle} \right ).
\end{equation}
For higgsino-like neutralino LSP, the annihilation cross section into
$W$-boson pair is estimated as \cite{Olive:1989jg}
\begin{equation}
	\langle \sigma v \rangle \simeq 
	\frac{\pi \alpha_2^2}{2} 
	\frac{m_{\mathrm{LSP}}^2}{(2m_{\mathrm{LSP}}^2-m_W^2)^2}
	\left ( 1- \frac{m_W^2}{m_{\mathrm{LSP}}^2} \right )^{3/2}
\end{equation}
and for wino-like neutralino LSP, it is given by~\cite{Moroi:1999zb}
\begin{equation}
	\langle \sigma v \rangle \simeq 
	8 \pi \alpha_2^2
	\frac{m_{\mathrm{LSP}}^2}{(2m_{\mathrm{LSP}}^2-m_W^2)^2}
	\left ( 1- \frac{m_W^2}{m_{\mathrm{LSP}}^2} \right )^{3/2}
\end{equation}
where $\alpha_2$ is the $SU(2)_L$ gauge coupling constant and $m_W$ is
the mass of $W$-boson.  We can see that the desired density of LSP can
be obtained for $m_{\rm LSP} \sim100$ GeV.  Thus both the dark matter
and baryon asymmetry of the universe can be explained even in the
presence of heavy modulus fields.

One may consider that these non-thermal LSPs from late-decaying
particles can have large free-streaming length $\lambda_{\rm FS}$
($\gtrsim 1$~Mpc ) and may become the warm dark matter
\cite{Moroi:1994rs,Borgani:1996ag,Kaplinghat:2005sy,Bringmann:2007ft}.  
Now we estimate the
free-streaming length of non-thermally produced LSPs assuming that they
contribute to the dark matter of the universe.  For simplicity, we
neglect the energy loss via the interaction between particles in thermal
bath.  Free streaming length of the LSP produced at $\tau _\chi$ is
given by~\cite{Cembranos:2005us}
\begin{equation}
   \lambda _{\mathrm{FS}}\sim 1.0 \mathrm{Mpc} ~u_d 
    \left ( \frac{\tau _\chi}{10^6\mathrm{sec}} \right )^{1/2}
	\left \{ 1+0.14 \ln \left [ \left ( 
	\frac{10^6 \mathrm{sec}}{\tau_\chi}\right )^{1/2}
	\frac{1}{u_d} \right ] \right \}
\end{equation}
where $u_d = \sqrt{m_\chi^2 -4 m_{\mathrm{LSP}}^2} /2m_{\mathrm{LSP}} $.
Using $\tau_\chi \sim (4\pi M_P^2/c m_\chi^3)$, we can rewrite it as
\begin{equation}
\begin{split}
	\lambda _{\mathrm{FS}}
	\sim &1.7 \times 10^{-2} \mathrm{Mpc} ~c^{-1/2} 
	\left ( \frac{100\mathrm{TeV}}{m_\chi} \right )^{1/2}
	\left ( \frac{1\mathrm{TeV}}{m_\mathrm{LSP}} \right ) \\
	& \times
	\left \{ 1 +0.07 \ln \left [  c^{1/2}
	\left ( \frac{m_\chi}{100\mathrm{TeV}} \right )^{1/2}
	\left ( \frac{m_\mathrm{LSP}}{1\mathrm{TeV}} \right ) 
	 \right ] \right \}.
\end{split}
\end{equation}
Thus, for $m_\chi \gtrsim 100$ TeV, free streaming length is much
smaller than 1 Mpc and non-thermal LSPs serve as the cold dark matter.
Actually, there exist non-negligible interactions of LSPs with
background particles.  It is expected that LSPs lose their energy and
momentum through those interactions and hence the non-thermal LSPs from
modulus decay unlikely take a role of warm dark matter
\cite{Kawasaki:1995cy,Hisano:2000dz}.

\subsection{Gravitinos from modulus decay}

If $m_\chi > 2m_{3/2}$, which is naturally realized in mirage-mediation
model, the modulus decay into two gravitinos is kinematically allowed.
In particular, it is found that such a decay mode generally has the
branching ratio as large as $O(0.01)$ \cite{Endo:2006zj,Dine:2006ii} and
the late-decay of gravitinos generated in this way may cause another
cosmological difficulty.  In our scenario these gravitinos do not upset
BBN, since they are also heavy enough to decay before the beginning of
BBN.  But LSPs emitted from the decay of those non-thermally produced
gravitinos may overclose the universe.

The decay of such non-thermal gravitinos does not release huge entropy,
because the energy density of the gravitino is two orders of magnitude
smaller than that of the ordinary radiation.  We denote the temperature
at the modulus decay and at the gravitino decay as $T_\chi$ and
$T_{3/2}$, respectively.  The branching ratios of moduli that decay into
ordinary radiation and two gravitinos are denoted as $B_r \sim O(1)$ and
$B_{3/2} $.  In these terms, the ratio of the energy density of
gravitino to radiation at the decay of gravitino is given by
\begin{equation}
   \frac{\rho_{3/2}}{\rho_r}
    =\bar \epsilon \frac{B_{3/2}}{B_r} 
\end{equation}
where
\begin{equation}
 \bar \epsilon = \left \{
	     \begin{array}{ll}
	      \dfrac{T_{\mathrm{NR}}}{T_{3/2}}~~~~~
	       &(T_{\mathrm{NR}}>T_{3/2} )\\
		1  ~~~~~&(T_{\mathrm{NR}}< T_{3/2})
	     \end{array}
	\right.
\end{equation}
and $T_{\mathrm{NR}} (= (m_{3/2}/m_\chi)T_\chi) $
denotes the temperature at which gravitinos become non-relativistic.
Since $B_{3/2} \ll B_r$, $\rho_{3/2}$ is not larger than $\rho_r$ for
$T_{\mathrm{NR}}< T_{3/2}$. When the gravitino becomes non-relativistic
before decay, we obtain
\begin{equation}
	\frac{\rho_{3/2}}{\rho_r}  = \frac{B_{3/2}}{B_r} 
	\frac{T_{\chi}}{T_{3/2}}  \frac{m_{3/2}}{m_\chi} \sim 
	\frac{B_{3/2}}{B_r} \left ( \frac{m_\chi}{m_{3/2}} \right )^{1/2}.
\end{equation}
This ratio does not exceed 1 in the parameter region we are interested
in, and baryon asymmetry is not diluted further by the gravitino decay.
The difficulty comes from the subsequent decay of gravitinos into LSPs.
The LSP abundance emitted from gravitino decay is also expressed by
Eq.~(\ref{nonthermalLSP}) after replacing $T_\chi$ with $T_{3/2}$.  But for
$m_{3/2}\sim 100$ TeV, $T_{3/2}$ is so small that the LSPs do not
annihilate and their density overcloses the universe.  To avoid this
difficulty, we require $m_{3/2} \gtrsim 10^3$ TeV.  But in such a case,
the mass of LSP becomes too large in anomaly-mediation or
mirage-mediation models and the overclosure problem of LSPs is not
cured.

Here we describe some ways to avoid the LSP overproduction problem
in such heavy moduli scenario with $m_\chi \gg m_{3/2}$.
One possible solution is to reduce $B_{3/2}$ so that the abundance of
gravitinos from moduli decay can be neglected.  Depending on the Kahler
potential and SUSY breaking sector, the branching ratio of modulus decay
into gravitinos may have the helicity suppression factor $\sim
(m_{3/2}/m_\chi)^2$ compared with other decay modes \cite{Dine:2006ii}.

Another is to introduce lighter $R$-odd particles other than MSSM
particles.  Axino, which appears in supersymmetric extension of the
axion models \cite{Kim:1986ax}, is one of the candidates
\cite{Rajagopal:1990yx}.  In such a case, the overproduced lightest
neutralino eventually decays into axinos and its abundance is reduced by
the factor $(m_{\tilde a}/m_\mathrm{LSP})$ where $m_{\tilde a}$ is the
axino mass.  If $m_{\tilde a}$ is sufficiently small \cite{Goto:1991gq},
the overproduction problem of neutralino LSP can be solved.  It should
be noticed that it may also open the decay mode of gravitino into axino
and axion \cite{Asaka:2000ew}, and this newly produced non-thermal
axions serve as the additional radiation energy density
\cite{Ichikawa:2007jv}, which speeds-up the Hubble expansion and changes
the result of BBN especially the { $^4$He } abundance.  In terms of the
effective number of neutrinos $N_\nu$, the success of BBN requires
$\Delta N_\nu \lesssim 1$ at the beginning of BBN \footnote{The recent
analysis of primordial {$^4$He } abundance favors non-standard value of
$N_\nu ( > 3) $ \cite{Izotov:2007ed}, but $\Delta N_\nu \sim 1$ is
disfavored. }.  But in our situation this is not a problem, since the
gravitino abundance is smaller than the radiation at the decay of
gravitinos and hence the axion abundance generated from the gravitino
decay is also smaller than the radiation energy density.  
In this scenario axinos 
should decay before BBN \cite{Covi:1999ty}.
Otherwise, decay products of the neutralino spoils BBN.
This requires the Peccei-Quinn scale 
$10^{10}$\,GeV$\lesssim F_{\rm PQ} \lesssim 10^{11}$\,GeV,
where the lower bound comes from the astrophysical consideration \cite{Raffelt:1990yz}
and the upper bound comes from the requirement the lifetime of the decaying neutralino
into axino should be shorter than 1 sec.
Thermally produced axinos \cite{Covi:2001nw} 
can contribute to the only small fraction of the energy density
of the universe because the final reheating temperature is very low.
The coherent oscillation of the axion is also diluted by the modulus decay
and has neglecting effects on cosmology for $F_{\rm PQ} \lesssim 10^{11}$\,GeV
\cite{Kawasaki:1995vt}. 

Finally, we mention the possibility that gravitinos are diluted by
entropy production after the modulus decay in the following subsection.
This is already built in the models of Q-ball dominant universe
(Sec.~\ref{sec:W=0}), as we will see.

\subsection{Dilution by Q-ball decay}

The abundance of gravitinos from modulus decay is expressed as
\begin{equation}
	Y_{3/2} = 2\frac{B_{3/2}}{B_r}\frac{3T_\chi}{4m_\chi}.
\end{equation}
Assuming no annihilation, the energy density of the LSP produced by the
decay of gravitinos is given by
\begin{equation}
	\Omega_{\mathrm{LSP}}h^2 \sim 2.3\times 10
	\sqrt c
	\left ( \frac{B_{3/2}/B_r}{0.01} \right )
	\left ( \frac{m_{\mathrm{LSP}}}{100\mathrm{GeV}} \right )
	\left ( \frac{m_\chi}{100 \mathrm{TeV}} \right )^{1/2}.
\end{equation}
Thus we need the dilution factor $\Delta \sim 10^{2}-10^{3}$ after the
production of gravitinos.  In fact, in the Q-ball dominant case
investigated in Sec.~\ref{sec:W=0}, the decay of Q-balls releases large
entropy and dilutes the gravitinos from modulus decay.  If we denote the
decay temperature of Q-balls as $T_Q$, the dilution factor is given by
\begin{equation}
	\Delta = \frac{T_\chi}{T_Q}
	 \left ( \frac{| \psi_0|}{|\chi_0|} \right )^2.
\end{equation}
Thus if $T_Q$ is slightly smaller than $T_\chi$, and the initial
amplitude of the modulus field $|\chi_0|$ is slightly smaller than that
of the Affleck-Dine field $|\psi_0|$, the required dilution factor is
obtained.  This is no more than the situation we encountered in
Sec.~\ref{sec:W=0}, and hence those Q-balls give the desired dilution
factor, solving the overproduction problem of LSPs from gravitino decay.
Note also that LSPs are also produced from decay of Q-balls.  Their
abundance is given by Eq.~(\ref{nonthermalLSP}) after replacing $T_\chi$
with $T_Q$.  As far as $T_Q \gtrsim 100$ MeV, LSPs can effectively
annihilate and their abundance becomes below or comparable to that of the
dark matter \cite{Fujii:2001xp,Fujii:2002kr}.

\begin{figure}[htbp]
 \begin{center}
  \includegraphics[width=0.65\linewidth]{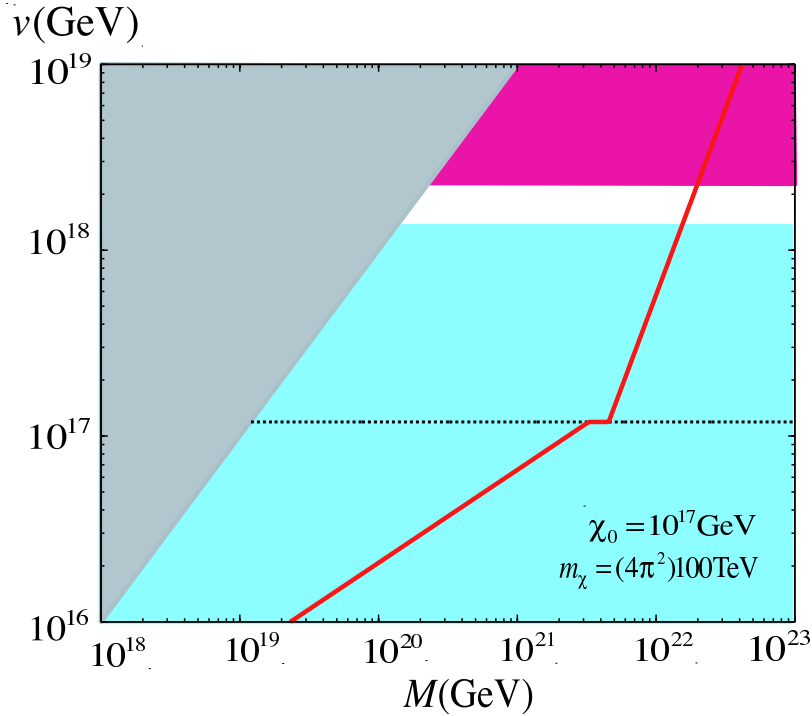} \caption{ Same as
  Fig.~\ref{fig:W=0_1}, but for $\chi_0=10^{17}$\,GeV.  We also show the
  constraint from LSP overproduction from gravitino decay for $m_\chi =
  4\pi^2 100$ TeV.  In the blue shaded region, the Q-ball decay can not dilute
  LSPs produced by the decay of moduli-induced gravitinos
  sufficiently. 
  In the purple shaded region, LSPs produced by Q-ball decay overclose the universe.
  } \label{fig:W=0_2}
 \end{center}
\end{figure}

In Fig.~\ref{fig:W=0_2}, the result with $|\chi_0|=10^{17}$\,GeV is shown.  It
can be seen that the wide parameter region which has been favored in
Sec.~\ref{sec:W=0} is excluded by the constraint from the overproduction
of LSPs both from the Q-ball decay (the purple shaded region) 
and gravitino decay (the blue shaded region).  We can see that
only in narrow parameter region for $M\sim 10^{22}$ GeV and $v \sim
M_P$, both the dark matter as non-thermal LSPs and baryon asymmetry can
be explained simultaneously.  
As the value of $\chi_0$ is reduced, the constraint is relaxed.
Although some degree of tunings to the
parameters especially the initial amplitude of the modulus and
Affleck-Dine field is required, this is a fully consistent cosmological
scenario in the presence of heavy moduli.


Finally we comment on the possibility where the similar dilution from
Q-ball decay is obtained in the model of Sec.~\ref{sec:earlyosc}.  In
order to realize this, we consider the situation where a flat direction
other than Affleck-Dine field responsible for baryogenesis
dominates the universe after the moduli decay.  In fact, $udd$ and $LLe$
directions can have the large field value simultaneously.  As described
in Sec.~\ref{sec:earlyosc}, $n=6~udd$ direction is used as the
Affleck-Dine field which create the appropriate baryon number.  On the other
hand, $LLe$ direction also has the large field value.  We assume for the
$LLe$ direction there do not exist non-renormalizable superpotentials
which lift the direction and we parametrize this direction as $\psi$
(Such a model was considered in Ref.~\cite{Ichikawa:2004pb}.).  Similar
to the case in Sec.~\ref{sec:W=0}, $\psi$ has the initial amplitude
$\psi_0$ of order $M_P$.
The late decay of Q-balls from $\psi$-condensate dilutes the gravitino
to the cosmologically safe value.  Although it also dilutes the baryon
asymmetry by the factor $\Delta \sim 10^{2}$, dilution of such amount of
baryon asymmetry is not so harmful, as can be seen from
Fig.~\ref{fig:m_chi=1e7}.

\section{Conclusions} \label{sec:conclusion}

Within the framework of fundamental theory such as supergravity or
superstring theory, there appear cosmologically harmful scalar fields
called moduli.  In anomaly-mediated SUSY breaking or mirage mediation
model, moduli are heavy and decay well before BBN starts, but the decay
process dilutes the preexisting baryon asymmetry.  We have shown that in
some models of Affleck-Dine baryogenesis mechanism, large amount of
baryon asymmetry can be generated and can survive the dilution from the
modulus decay. Successful baryogenesis requires high reheating
temperature from inflaton, $T_R \gtrsim 10^{10}$ GeV, and high effective
cutoff scale, $M \gtrsim 10^{19}$ GeV for early oscillation models.
Such a high reheating temperature is naturally realized in chaotic
inflation models \cite{Linde:1983gd}.  Gauged $U(1)_{B-L}$ models also
work for some parameter regions.  We also investigated the gauged
$U(1)_{B-L}$ model without superpotentials which lift the flat
direction.  The favored parameter region is also found in this type of
model.  Other baryogenesis mechanisms such as thermal leptogenesis
\cite{Fukugita:1986hr} and electroweak baryogenesis
\cite{Trodden:1998ym} do not work, since produced baryon aynmmetry is
not so large as to survive the dilution.

Aside from baryon asymmetry, dark matter of the universe can also be
explained by the non-thermal LSPs from the decay of moduli.  The final
reheating temperature is determined by the decay of moduli and it is
predicted as from a few MeV to 1 GeV for $100$ TeV $\lesssim m_\chi
\lesssim (4\pi^2) 100$ TeV.  Hence the standard cosmological scenario
below a few MeV should not be changed.  One subtlety arises when we
consider the gravitino production from decay of the heavy moduli if the
modulus mass $m_\chi$ is larger than two times the gravitino mass
$m_{3/2}$.  If the branching ratio of modulus decay into two gravitinos
is not suppressed, we encounter the another cosmological problem,
i.e., overproduction of neutralino LSPs from the subsequent decay
of gravitinos.  In the Q-ball dominant scenario in Sec.~\ref{sec:W=0},
Q-ball decay dilutes the gravitino and the problem can be solved by
choosing the initial amplitude of the modulus and Affleck-Dine field as
$|\chi_0|\lesssim |\psi_0| \sim M_P$.  Besides Q-ball dominant scenario,
there are a couple of ways to avoid this difficulty. One is controlling
the SUSY breaking sector to suppress the branching ratio into
gravitinos, and another is to introduce the axinos.  The other solution
is to invoke another flat direction condensate into large Q-balls.  The
late-decay of Q-balls dilutes the gravitino abundance, and also such a
Q-ball decay itself can provide the non-thermal origin of the dark
matter, similar to the Q-ball dominant model.  In any way, our scenario
provides the realistic cosmological scenario in the presence of modulus
fields and may have phenomenologically interesting implication to future
collider experiments and direct or indirect detection of the dark
matter.

\begin{acknowledgments}

We are grateful to M.~Endo and F.~Takahashi for useful comments.
K.~N.~would like to thank the Japan Society for the Promotion of Science for financial support.
This work was supported in part by the Grant-in-Aid for Scientific Research
from the Ministry of Education, Science, Sports, and Culture of Japan,
No. 18540254 and No 14102004 (M.K.).  This work was also supported
in part by JSPS-AF Japan-Finland Bilateral Core Program (M.K.)

\end{acknowledgments}


\end{document}